\input{aipcheck}

\documentclass[
    ,final            
  ]
  {aipproc}
\usepackage{sidecap,graphicx,boxedminipage,epsfig,wrapfig,floatflt,shadow}

\layoutstyle{6x9}

\begin{document}

\title{Problem Solving and Learning}

\classification{01.40.Fk,01.40.G-,01.40.gb}
\keywords      {Cognitive Science, Problem Solving, interactive tutorials, introductory physics}

\author{Chandralekha Singh}{
  address={Department of Physics and Astronomy, University of Pittsburgh, Pittsburgh, Pennsylvania, 15260}
}

\begin{abstract}
One finding of cognitive research is that people do not automatically
acquire usable knowledge by spending lots of time on task. Because students'
knowledge hierarchy is more fragmented, "knowledge chunks" are smaller than
those of experts. The limited capacity of short term memory makes the
cognitive load high during problem solving tasks, leaving few cognitive
resources available for meta-cognition.
The abstract nature of the laws of physics and the chain of reasoning
required to draw meaningful inferences makes these issues critical.
In order to help students, it is crucial to consider the difficulty of a problem from the perspective of students.
We are developing and evaluating interactive problem-solving tutorials to help students in the introductory physics courses learn
effective problem-solving strategies while solidifying physics concepts.
The self-paced tutorials can provide guidance and support for a variety of problem solving techniques, and opportunity for knowledge
and skill acquisition.
\end{abstract}

\maketitle


\vspace*{-.15in}
\section{Cognitive Research and Problem Solving}
\vspace*{-.05in}

Cognitive research deals with how people learn and solve problems~\cite{nrc1,joe}. At a coarse-grained level, there are three
components of cognitive research: how do people acquire knowledge, how do they organize and retain the knowledge in memory (brain)
and how do they retrieve this knowledge from memory in appropriate situations including to solve problems. These three components
are strongly coupled, e.g., how knowledge was organized and retained in memory during acquisition will determine how effectively it can be retrieved
in different situations to solve problems. We can define problem solving as any purposeful activity where one
must devise and perform a sequence of steps to achieve a set goal when presented with a novel situation.
A problem can be quantitative or conceptual in nature.

Using the findings of cognitive research, human memory can be broadly divided into two components: the working memory or the short term memory
(STM) and the long term memory (LTM). The long term memory is where the prior knowledge is stored.
Appropriate connections between prior knowledge in LTM and new knowledge that is being acquired at a given time can help an individual organize
his/her knowledge hierarchically. Such hierarchical organization can provide indexing of knowledge where more fundamental concepts
are at the top of the hierarchy and the ancillary concepts are below them. Similar to an index in a book, such indexing of knowledge
in memory can be useful for accessing relevant knowledge while solving problems in diverse situations. It can also be useful for
inferential recall when specific details may not be remembered. 

The working memory or STM is where information presented
to an individual is processed. It is the conscious system that receives input from memory buffers associated with various sensory
systems and can also receive input from the LTM. Conscious human thought and problem solving involves rearranging and synthesizing
ideas in STM using input from the sensory systems and LTM.

One of the major initial findings of the cognitive revolution is related to Miller's magic numbers 7$\pm$2 (5 to 9), i.e., how much information
can STM hold at one time.~\cite{miller} Miller's research found that STM can only hold 5 to 9 pieces of information regardless of the IQ of
an individual. Here is an easy way to illustrate this. If an individual is asked to memorize the following
sequence of 25 numbers and letters in that order after staring at it for 30 seconds, it is a difficult task:
6829-1835-47DR-LPCF-OGB-TWC-PVN.
An individual typically only remembers between 5 to 9 things in this case. However, later research shows that people can extend the limits of
their working memory by organizing disparate bits of information into chunks or patterns.~\cite{chunk} Using chunks, STM can evoke from
LTM, highly complex information. An easy way to illustrate it is by asking an individual to memorize the following sequence of
25 numbers and letters: 1492-1776-1865-1945-AOL-IBM-USA. This task is much easier if one recognizes that each of the four digit number
is an important year in history and each of the three letters grouped together is a familiar acronym. Thus, an individual only has
to remember 7 separate chunks rather than 25 disparate bits. This chunking mechanism is supported by research in knowledge rich fields
such as chess and physics where experts in a field have well organized knowledge.~\cite{chess} For example, research shows that if experts in
chess are shown a very good chess board that corresponds to the game of a world-class chess player, they are able to assemble the board
after it is disassembled because they are able to chunk the information on the board and remember the position of one piece with respect
to another. If chess novices are shown the same board, they are only able to retrieve 5-9 pieces after it is jumbled up
because they are not able to chunk large pieces of information present on the chess board. On the other hand, both chess experts and novices
are poor at assembling a board on which the chess pieces are randomly placed before it was jumbled up. In this latter case, chess experts
are unable to chunk the random information due to lack of pattern.

A crucial difference between expert and novice problem solving is the manner in which knowledge is represented in their memory
and the way it is retrieved to solve problems. Experts in a field have well organized knowledge. They have large chunks of
``compiled" knowledge in LTM and several pieces of knowledge can be accessed together as a chunk~\cite{automatic}.
For example, for an expert in physics, vector addition, vector subtraction, displacement,
velocity, speed, acceleration, force etc. can be accessed as one chunk while solving problems while they can be
seven separate pieces of information for beginning students. If a problem involves all of these concepts, it
may cause a cognitive overload if students' STM can only hold 5 or 6 pieces of information.
Experts are comfortable going between different knowledge representations, e.g., verbal, diagrammatic/pictorial, tabular etc. and
employ representations that make problem solving easier.~\cite{rep} Experts categorize problems based upon deep features unlike novices who
can get distracted by context dependent features. For example, when physics professors and introductory physics students are asked
to group together problems based upon similarity of solution, professors group them based upon physics concepts while students
can choose categories that are dependent on contexts such as ramp problems, pulley problems, spring problems etc~\cite{chi,hardiman,reif2,larkin}.

Of course, an important goal of most physics courses is to help students develop expertise in problem solving and improve
their reasoning skills. In order to help students, instructors must realize that the cognitive load,
which is the amount of mental resources needed to solve a problem, is subjective~\cite{cogload}. The complexity of a problem not only depends
on its inherent complexity but also on the expertise, experience and intuition of an individual~\cite{intuition}. It has been said that
problems are either ``impossible" or ``trivial". A ballistic pendulum problem that may be trivial for a physics professor may
be very difficult for a beginning student~\cite{rosengrant}.
Cognitive load is higher when the context is abstract as opposed to concrete. The following Wason tasks~\cite{wason} are examples of
abstract and concrete problems which are conceptually similar, but the abstract problem turns out to be cognitively more demanding.
\begin{itemize}
\item \underline{Abstract Task:} You will lose your job unless you enforce the following rule:
``If a person is rated K, then his/her document must be marked with a 3". \\
Each card on the table for a person has a letter on one side and a number on the other side.
Indicate only the card(s) shown in Figure 1 that
you definitely need to turn over to see if the document of any of these people violates this rule. \\

\begin{center}
\begin{figure}[h!]
\epsfig{file=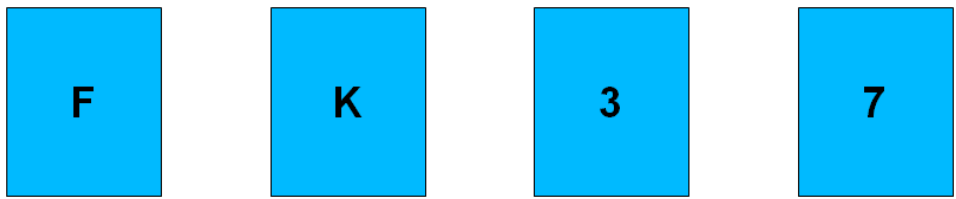,height=1.in}
\vspace*{-.3in}
\caption{Figure for Wason Task in abstract context~\cite{wason}}
\end{figure}
\end{center}

\item \underline{Concrete Task:} You are serving behind the bar of a city centre pub and will lose your job unless you enforce the following rule:
``If a person is drinking beer, then he/she must be over 18 years old". \\
Each person has a card on the table which has his/her age on one side and the name of his/her drink on the other side.
Indicate only the card(s) shown in Figure 2 that you definitely need to turn over to see if any of these people are breaking this rule.

\begin{center}
\begin{figure}[h!]
\epsfig{file=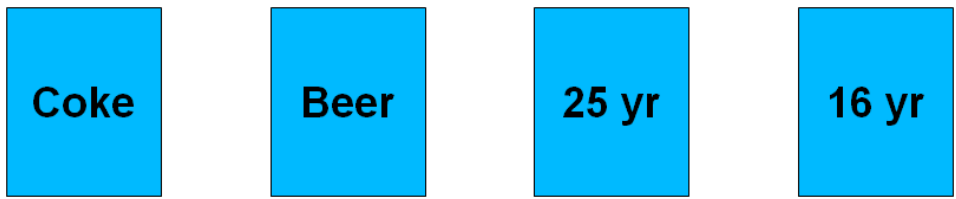,height=1.in}
\vspace*{-.3in}
\caption{Figure for Wason Task in concrete context~\cite{wason}}
\end{figure}
\end{center}

\end{itemize}

The correct answer for the abstract case is that you must turn the cards with K and 7 (to make sure that there is no K on the other side).
Please note that the logic presented in the task is one sided in that it is ok for a document with a 3 to have anything on the other side.
The correct answer for the concrete case is ``beer" and ``16 years old", and it is much easier to identify these correct answers
than the correct answers for the abstract case.
A major reason for why the cognitive load is high during problem solving in physics is because the laws of physics are abstract.
It is important to realize that it is not easy to internalize them unless concrete contexts are provided to the students.
Another difficulty is that, once the instructor has built an intuition about a problem, it may not appear difficult to him/her even
if it is abstract. In such situations the instructor may overlook the cognitive complexity of the problem for a beginning student
unless the instructor puts himself/herself in the students' shoes.

An important lesson from cognitive research is that new knowledge that an individual acquires builds on prior knowledge. This idea
is commensurate with Piaget's notion of ``optimal mismatch"~\cite{piaget} or Vygotsky's idea of ``zone of proximal development" (ZPD)~\cite{vygotsky}. ZPD is the
zone defined by what a student can do on his/her own vs. with the help of a guide who is familiar with the student's initial knowledge
and targets instruction somewhat above it continuously for effective learning. This is analogous to the impedance matching of a transformer
in which the power transfer can be maximized if the input and output impedances are matched. Another analogy is with light passing through
two polarizers placed perpendicular to each other vs. having several polarizers stacked one after another where the transmission axes of adjacent polarizers are
slightly different from each other. In the first case of crossed polarizer, no light passes through whereas in the second case most of the
light passes through if the angle $\theta$ between the transmission axes of the adjacent polarizers is small enough. Similarly, if the
instruction is targeted significantly above students' prior knowledge, learning won't be meaningful and even if the students
make an effort to store some hap-hazardous information in their brain till the final exam, it will get ``shampooed out" soon after that.
On the other hand, if the instructional design takes into account students' initial knowledge and builds on it, learning will be meaningful.

Another important lesson from cognitive research is that students must construct their own understanding. This implies that
we should give students an opportunity to reconstruct, extend, and organize their knowledge. Such opportunities will come from
ensuring that the students are actively engaged in the learning process and take advantage of their own knowledge resources and
also benefit from interactions with their peers.

\vspace*{-.15in}
\section{Computer-based Interactive Tutorials}
\vspace*{-.05in}

We now describe computer-based interactive
problem solving tutorials that we have been developing that build on introductory physics students' prior knowledge and keep them
actively engaged in the learning process. The tutorials combine
quantitative and conceptual problem solving. They focus on helping students develop a functional understanding of physics while learning
useful skills~\cite{singh}.
It is worthwhile thinking about why quantitative problem solving alone often fails to help most students extend and organize their physics
knowledge. Without guidance, most students do not exploit the problem solving opportunity to
reflect upon what they have actually learned and build a more robust knowledge structure.
If only quantitative problems are asked, students often view them as ``plug-and-chug" exercises, while
conceptual problems alone are often viewed as guessing tasks with little connection to physics content.
The interactive tutorials we have been developing combine quantitative and conceptual problem solving and provide guidance and support for
knowledge and skill acquisition.
They provide a structured approach to problem solving and promote active engagement while helping students develop self reliance.
Other computer-based tutorials are also being developed~\cite{tutor}.
Our tutorials are unique in that they focus on helping students learn effective problem solving strategies and
the conceptual questions are developed based upon the common difficulties found via research on students' difficulties in learning
a particular topic in physics.

\vspace*{-.15in}
\subsection{Development of Problem Solving Skills in Introductory Physics}
\vspace*{-.05in}

A major goal of an introductory physics course for science and engineering majors
is to enable students to develop complex reasoning and problem solving skills
to explain and predict diverse phenomena in everyday experience.
However, numerous studies show that students do not acquire these skills from a {\it traditional} course~\cite{chi,hardiman,reif2,hake}.
The problem can partly be attributed to the fact that
the kind of reasoning that is usually learned and employed in everyday life is not systematic or rigorous.
Although such hap-hazardous reasoning may have little measurable negative consequences in an individual's personal life, it is insufficient to
deal with the complex chain of reasoning that is required in rigorous scientific field such as physics~\cite{reif2}.

Educational research suggests that many introductory physics students solve problems using superficial clues and cues,
applying concepts at random without thinking whether they are applicable or not~\cite{chi,hardiman,reif2}.
Also, most traditional courses do not {\it explicitly} teach students effective problem solving strategies.
Rather, they may reward inferior problem solving strategies in which many students engage.
Instructors often implicitly assume that students know that the analysis, planning,
evaluation, and reflection phases of problem solving are as important as the implementation phase.
Consequently, they may not discuss these strategies explicitly while solving problems during the lecture.
There is no mechanism in place to ensure that students make a conscious effort to
interpret the concepts, make qualitative inferences from the quantitative problem solving tasks,
or relate the new concepts to their prior knowledge.

In order to develop scientific reasoning by solving quantitative problems, students must
learn to exploit problem solving as an opportunity for knowledge and skill acquisition.
Thus, students should not treat quantitative problem solving merely as a mathematical exercise but as a learning opportunity and
they should engage in effective problem solving strategies.

\vspace*{-.15in}
\subsection{Effective Problem Solving Strategies}
\vspace*{-.05in}

Effective problem solving begins with a conceptual analysis of the problem, followed by planning of the problem solution,
implementation and evaluation of the plan, and last but not least reflection upon the problem solving process.
As the complexity of a physics problem increases, it becomes increasingly important to employ a systematic approach.
In the qualitative or conceptual analysis stage, a student should draw a picture or a diagram and get a visual
understanding of the problem. At this stage, a student should convert the problem to a representation that makes 
further analysis easier. After getting some sense of the
situation, labeling all known and unknown numerical quantities is helpful in making reasonable physical assumptions.
Making predictions about the solution is useful at this level of analysis and it can help to structure the decision making at
the next stage.
The prediction made at this stage can be compared with the problem solution in the reflection phase and can help repair, extend and organize
the student's knowledge structure.
Planning or decision making about the applicable physics principles is the next problem solving heuristic.
This is the stage where the student brings everything together to come up with a reasonable solution. If the student performed good
qualitative analysis and planning, the implementation of the plan becomes easy if the student possesses the necessary
algebraic manipulation and mathematical skills.

After implementation of the plan, a student must evaluate his/her solution, e.g., by checking the dimension or the order of magnitude,
or by checking whether the initial prediction made during the initial analysis stage matches the actual solution.
One can also ask whether the solution is sensible and, possibly, consistent with experience.
The reflection phase of problem solving is critical for learning and developing expertise.
Research indicates that this is one of the most neglected phase of problem solving~\cite{chi,hardiman,reif2}.
Without guidance, once a student has an answer, he/she typically moves on to the next problem.
At the reflection stage, the problem solver must try to distill what he or she has learned from solving the problem. This
stage of problem solving should be used as an opportunity for reflecting upon why a particular principle of physics is
applicable to the problem at hand and how one can determine in the future that the same principle should be applicable even if
the problem has a new context.

\vspace*{-.15in}
\subsection{Description of the Tutorials}
\vspace*{-.05in}

The development of the computer-based tutorials to help students learn effective problem solving strategies
is guided by a learning paradigm which involves three essential components: modeling, coaching, and weaning~\cite{cog}.
In this approach, ``modeling" means that the instructor
demonstrates and exemplifies the skills that students should learn (e.g., how to solve physics problems systematically).
``Coaching" means providing students opportunity, guidance and
practice so that they are actively engaged in learning the skills necessary for good performance. ``Weaning" means
reducing the support and feedback gradually so as to help students develop self-reliance.

Each of the tutorials starts with an overarching problem which is quantitative in nature.
Before using a tutorial, students use a pre-tutorial worksheet
which divides each quantitative problem given to them into different stages involved in problem solving.
For example, in the conceptual analysis stage of problem solving, the worksheet explicitly asks students to draw
a diagram, write down the given physical quantities, determine the target quantity, and predict some features of the solution. After attempting the problem on the worksheet to the best of their ability, students access the tutorial
on the computer (or use a paper version for evaluation purposes as discussed in the evaluation section below).
The tutorial divides an overarching problem into several sub-problems, which are research-guided conceptual
multiple-choice questions
related to each stage of problem solving. The alternative choices in these multiple-choice questions elicit common difficulties students
have with relevant concepts as determined by research in physics education. Incorrect responses direct students to appropriate
help sessions where students have the choice of video, audio or only written help with suitable explanations, diagrams, and equations.
Correct responses to the multiple-choice questions give students a choice of either advancing to the next sub-problem or
directs them to the help session with the reasoning and explanation as to why the alternative choices are incorrect.
While some reasonings are problem-specific, others focus on more general ideas.

After students work on the implementation and assessment phase sub-problems posed in the multiple-choice format, they answer reflection sub-problems.
These sub-problems focus on helping students reflect upon what they have learned and apply the concepts learned in different contexts.
If students have difficulty answering these sub-problems, the tutorial provides further help and feedback.
Thus, the tutorials not only model or exemplify a systematic approach to problem solving, they also engage students actively in the learning process
and provide feedback and guidance based upon their need.

Each tutorial problem is matched with other problems (which we call paired problems) that use similar physics principles but which are somewhat
different in context. Students can be given these paired problems as quizzes so that they learn to de-contextualize the problem
solving approach and concepts learned from the tutorial.
The paired problems play an important role in the weaning part of the learning model and ensure that
students develop self-reliance and are able to solve problems based upon the same principle without help.
These paired problems can also be assigned as homework problems and instructors can inform students that they can use the tutorials as
a self-paced study tool if they have difficulty in solving the paired problems assigned as homework related to a particular topic.

\begin{center}
\begin{figure}[h!]
\epsfig{file=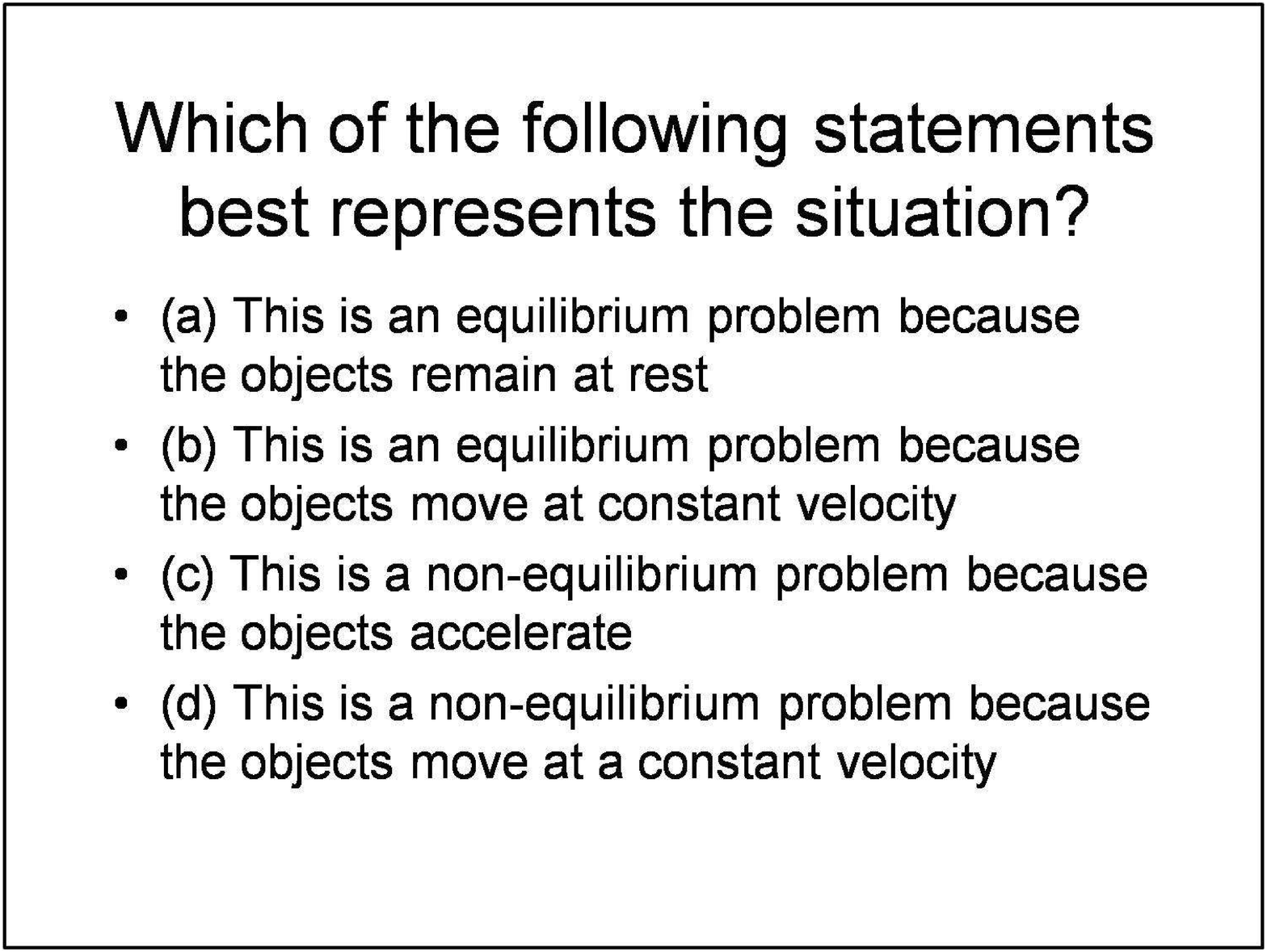,height=2.05in}
\epsfig{file=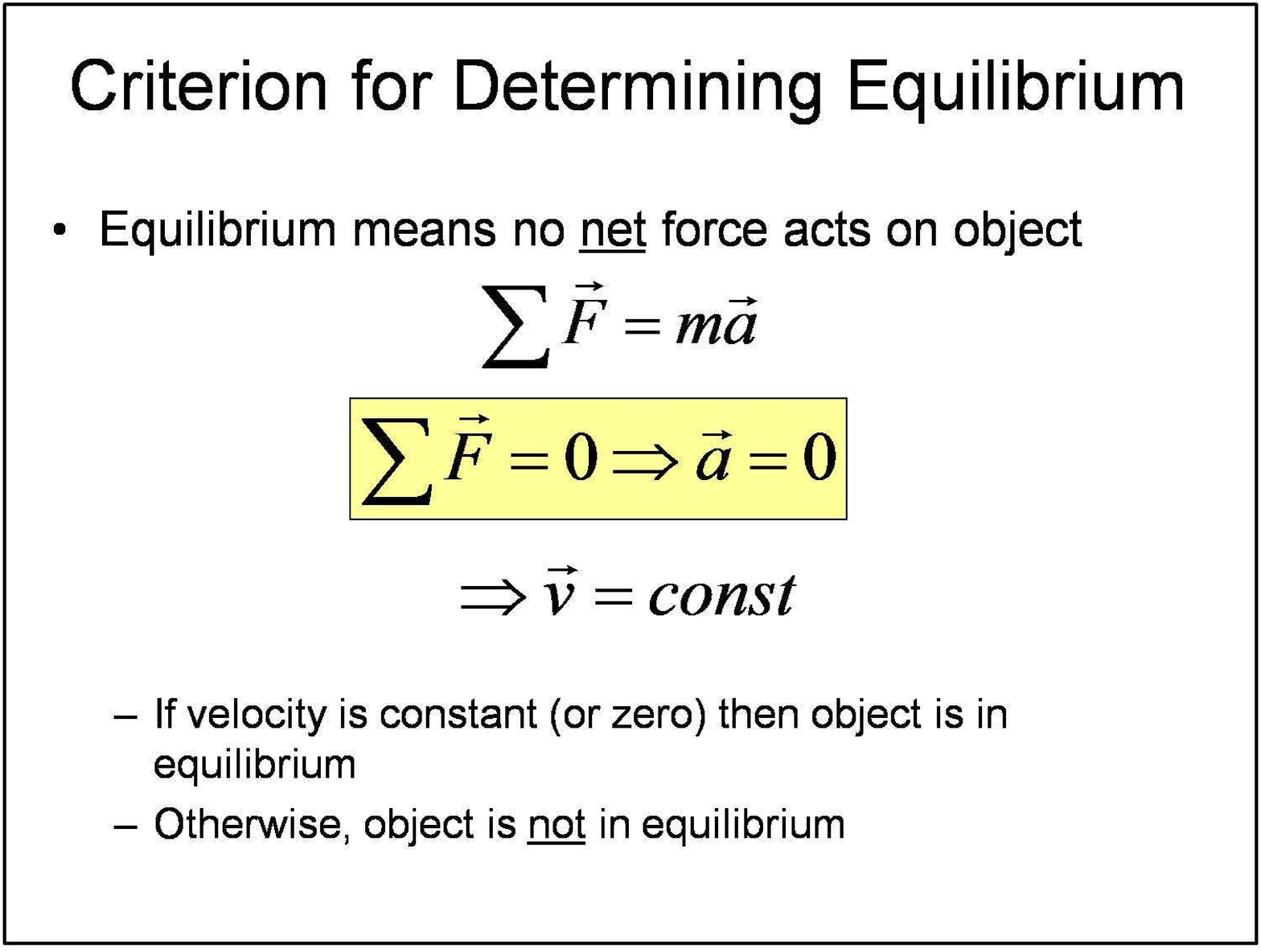,height=2.05in}
\caption{An example of a multiple-choice question and a related help screen. If students click on an incorrect choice, they are directed to the
help screen. Those choosing the correct option for the multiple-choice question can either advance to the next multiple-choice question
or can find out why the other choices are incorrect.}
\end{figure}
\end{center}

We have developed computer-based tutorials related to introductory mechanics, electricity, and magnetism. Topics in mechanics include linear
and rotational kinematics, Newton's laws, work and energy, and momentum. Topics in electricity and magnetism include Coulomb's law,
Gauss's law, potential and potential energy, motion of charged particles in an electric field, motion of charged particles
in a magnetic field, Faraday's law, and Lenz's law.

\begin{center}
\begin{figure}[h!]
\epsfig{file=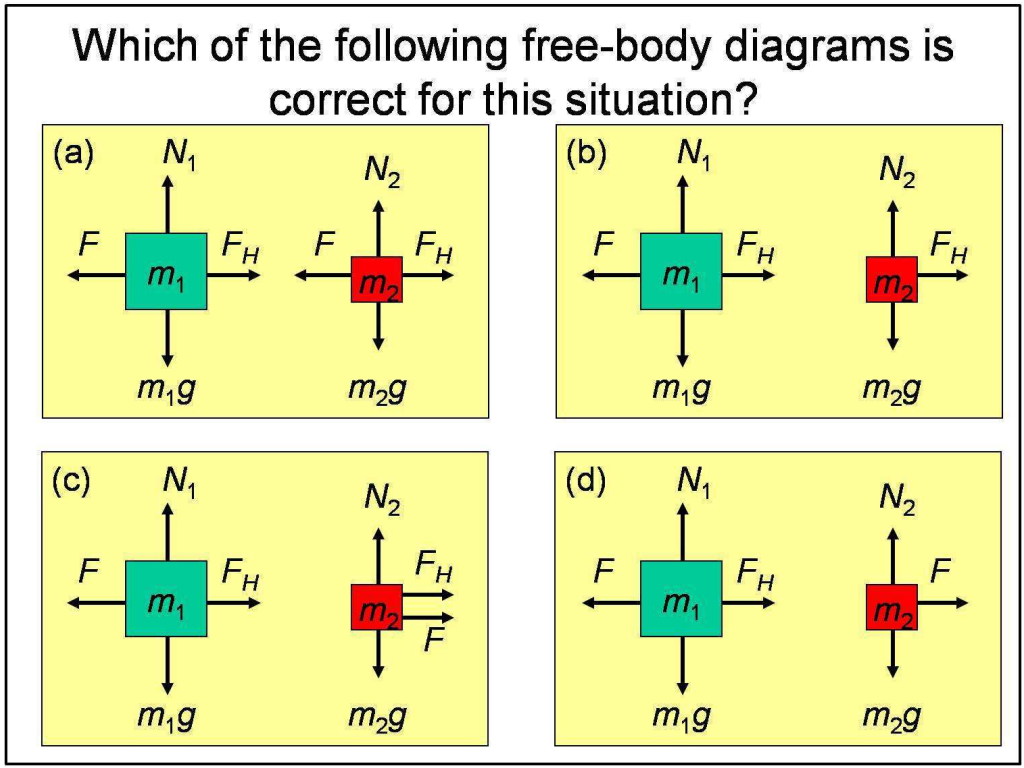,height=4.0in}
\vspace*{-.3in}
\caption{An example of a multiple-choice question related to the free body diagram}
\end{figure}
\end{center}


\begin{center}
\begin{figure}[h!]
\epsfig{file=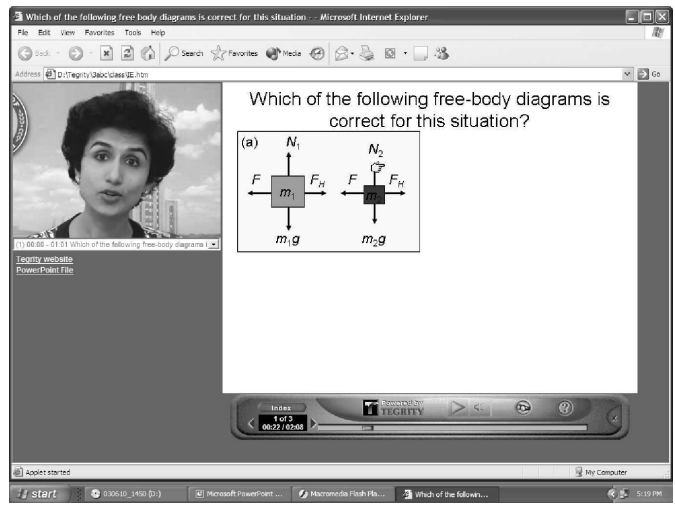,height=2.6in}
\vspace*{-.1in}
\caption{A screen capture of a help screen in the version of help in which an instructor guides students through the help session while
a cursor (hand shown near $N_2$ in the figure) moves and points to the relevant section of the power point screen.}
\end{figure}
\end{center}

\begin{center}
\begin{figure}[h!]
\epsfig{file=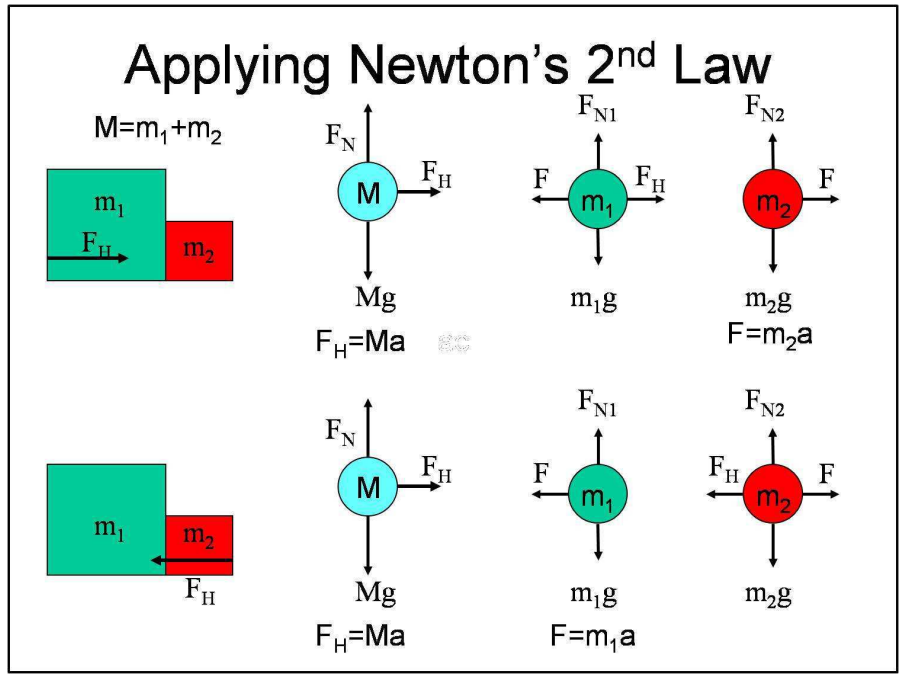,height=3.5in}
\vspace*{-.2in}
\caption{A screen capture of a help session for a reflection question in the tutorial.}
\end{figure}
\end{center}

\vspace*{-.25in}
Figures 3-6 show screen captures from a computer-based tutorial which starts with a quantitative problem in which two blocks with
masses $m_1$ and $m_2$ are in contact on a frictionless horizontal surface and a horizontal force $F_H$ is applied to the block
with mass $m_1$. Students are asked to find the magnitude of force exerted by the block with mass $m_2$ on $m_1$.
We have found that this problem is sufficiently
challenging for students in both algebra and calculus-based introductory physics courses that most students are unable to solve it without
help. In the tutorial, the quantitative problem is broken
down into several conceptual problems in the multiple-choice format that students have to answer. For example,
one of the conceptual questions related to the initial analysis of the problem is shown in Figure 3
along with a screen capture of a help session that a student is directed to if he/she chooses an incorrect response.
Figure 4 is a multiple-choice question about the free body diagram and figure 5 is a screen capture of a help screen in which an
instructor explains relevant concepts to the students related to difficulty with the question asked in Figure 4.
Figure 6 is a help screen related to a reflection question in which students are asked about the force exerted by the block of mass
$m_1$ on $m_2$ if the force of the hand $F_H$ was applied to the block of mass $m_2$ in the opposite direction (instead of being applied
to the block of mass $m_1$ as in the tutorial).

\vspace*{-.15in}
\subsection{Case-Study for Evaluating the Computer-based Tutorials}
\vspace*{-.05in}

Below, we describe a case-study to evaluate the tutorials. In one case study, we compared
three different groups who were given different aid tools:
\begin{itemize}
\item Group (1) consists of students who used the tutorials as aid tool.
\item Group (2) consists of students who were given the solved solutions for the tutorial problems which were
similar to the solutions in the textbook's solutions manual.
However, the solutions were not broken down into the multiple-choice questions with alternative choices targeting common misconceptions
as was done in the tutorials.
\item Group (3) consists of students who were given the textbook sections that dealt with the relevant concepts as the aid tool
and were asked to brush up on the material for a quiz on a related topic.
\end{itemize}
Fifteen students were recruited and divided into two pools based upon their prior knowledge. Then, the students
from each of these pools were randomly assigned to one of the three groups discussed above.

During the interview session, students in each group initially answered a pre-questionnaire
to determine their level of preparation, their views about problem solving in physics, and their perception of physics instruction.
It was interesting to note that a majority of the students (regardless of the group to which they belonged) disagreed with the statement ``When confronted
with a physics problem, I first spend a reasonable amount of time planning how to solve the problem before actually solving it".
Half of the students also agreed with the statement ``Physics problem solving is all about matching given quantities in the problem
to the formula in the book". Half of the students thought that the pace of their introductory physics courses was very fast.
After this pre-questionnaire, all students were given the following problem on a worksheet and were asked to solve it to the best of their
ability before using their aid tools:
\begin{itemize}
\item {\it An insulating sphere of radius $b$ has a spherical cavity of radius $a$ located within its volume and centered a distance $R$
from the center of the sphere. A cross section of the sphere is shown in Figure 7.
The solid part of the insulating sphere has a uniform volume charge density $\rho$.
Find the electric field  $\vec E$ at a point inside the cavity.}
\end{itemize}

\begin{center}
\begin{figure}[h!]
\epsfig{file=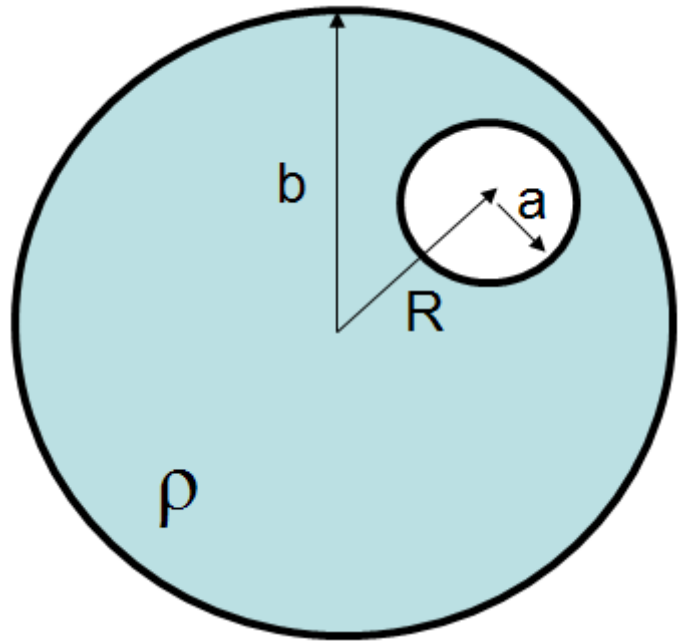,height=3.4in}
\caption{Figure associated with a tutorial problem based upon Gauss's law.}
\end{figure}
\end{center}

All students identified the correct principle to use (Gauss's law)
when asked to solve the problem to the best of their ability on the worksheet.
The above problem is challenging enough that none of the
interviewed students could solve it without help. Most students initially thought that the problem was relatively easy.
Except for one student who came up with a different incorrect answer, the rest of the
students came up with the same incorrect answer: the electric field is zero everywhere inside the cavity.
All of them invoked Gauss's law, which states that the total electric flux through a closed surface is equal to
the net charge inside the surface divided by $\epsilon_0$.
Their reasoning for zero electric field in the cavity was based on the incorrect interpretation that whenever the
electric flux through a closed surface is zero, the electric field at every point inside the surface must be zero too regardless
of whether there was a symmetric charge distribution to justify such claims.
Thus, students drew a Gaussian sphere inside the hole and concluded that the electric field must be zero everywhere inside
since the enclosed charge is zero. Students ignored the asymmetric charge distribution surrounding the cavity,
which is a common difficulty.~\cite{gauss}
Among many of the difficulties the interviewed students faced, one difficulty was
not recognizing that the charge distribution was not symmetric enough and therefore the net electric field at a point inside the
cavity cannot be zero.
When asked to show why the electric field should be zero everywhere inside, most students drew a spherical Gaussian surface inside
the hole, wrote down Gauss's law in the integral form and pulled out the electric field from inside the integral.
Interviews show that many students believed that the electric field can always be pulled out from the surface integral regardless of
the symmetry of the charge distribution. When pressed
harder about why the electric field should be equal everywhere on the Gaussian surface and why it
can be pulled out of the integral, some students noted that one should not worry about this issue at least in this case since the
zero charge enclosed implies zero electric field everywhere inside anyway.
Their convoluted reasoning showed that many students have difficulty in organizing different knowledge resources
and applying them at the same time.

After the students tried to solve the problem on the worksheet on their own to the best of their ability,
students in Group 1 were given the corresponding tutorial to work on, those in Group 2 were given a textbook-style solution for the problem
(similar to the solutions in a textbook solution manual), and those in Group 3 were given the section in the textbook,
University Physics by Young and Freedman, which deals with this topic.
Each student used his/her respective aid tool for the same amount of time (20 minutes). All students were told that they would
have to solve a paired problem involving similar concepts after help from the tools they were provided (tutorial, textbook-style
solution, relevant chapter from textbook). All students were informed that aid tool was only for learning the material and
could not be used while working on the paired problem they will be given later.

The paired problem that followed tested whether students could transfer relevant knowledge acquired from the aid tools to the
paired problem.~\cite{transfer} For example, for the Gauss's law problem discussed above, the paired problem was similar to
the tutorial problem but was for an infinite solid insulating cylinder with a uniform volume charge and an asymmetric cylindrical
cavity inside it. We used a rubric to grade students. The average performance of the tutorial group was approximately $85\%$.
All of them made the correct assumption that the electric field is not zero inside the cavity due to the asymmetry
of the charge distribution and explained how Gauss's law cannot be used in such cases to conclude that the electric
field is zero in the cavity.
During the interviews, these students were able to explain verbally their thought processes and how they solved the problem to find the
electric field. By analyzing the students' thought processes during the interviews, it appears that
the pattern of reasoning employed by these students was significantly better on an average than the reasoning of students from
the other two groups.

The other two groups didn't show as much improvement as the tutorial group when graded on a rubric after using the aids.
Between Groups 2 and 3, the students who used the textbook-style solution as a guide did better on the paired problem than
those who used the relevant textbook section.
The average performance of Group 2 was approximately $60\%$ and Group 3 was less than $30\%$.
Students who used the textbook-style solution still had difficulties in solving the paired problem and
four of them did not solve the entire problem correctly.
The solution of the problem involves breaking the problem into subproblems each of which
has a spherical symmetry and can be solved by known methods using Gauss's law. Most of the students in the second
group (those who were given a solution of the type given in solutions manual) realized that they had to combine or
superpose two electric fields. However, their most common difficulty was in using vectors in order to relate the final
solution to the solutions to the two subproblems which have a spherical symmetry.
The textbook-type solution showed them that calculating the
electric field at a point in the cavity involved subtracting from the electric field due to the full insulating sphere as though there was no
cavity, the electric field due to an insulating sphere of the size of the cavity.
From the surveys given after the paired problem, some students mentioned that the textbook-style solutions
didn't explain in words the steps used thoroughly.
Since the misconceptions students had at the beginning were not explicitly targeted by asking explicit questions in the textbook-type solution
(as was done explicitly in the multiple-choice questions which were part of the tutorials),
students did not transfer relevant knowledge from the solved example to the paired problem as well as the tutorial group did.

All of the students who made use of the textbook as an aid for learning (rather than the tutorial or the solved example) did poorly on the
paired problem. This finding is consistent with another study that shows that unless introductory physics students have seen solved examples.~\cite{iso2} Students in this group realized that the problem solution
could not be that the magnitude of the electric field inside the cavity is $\vert \vec E \vert=0$ because otherwise they would not be given 20 minutes to browse over the section of the textbook trying
to formulate a solution. But the responses they provided after browsing over the book were often difficult to understand and
dimensionally incorrect.
When asked to explain what they had done, students noted that they were not very sure about how to solve the problem. They
added that the relevant section of the textbook did not help because it did not have a solved example exactly like the problem
that was asked.


\vspace*{-.15in}
\section{Summary}
\vspace*{-.05in}

People do not automatically acquire usable knowledge by spending lots of time on task. Limited capacity of STM can make
cognitive load high for beginning students. For learning to be meaningful, students should be
actively engaged in the learning process. Moreover, it is important to consider the difficulty of a problem from students' perspective
and build on their prior knowledge.
We are developing computer-based interactive tutorials for introductory mechanics and electricity and magnetism
that are suited for a wide variety of students.
The self-paced tutorials combine quantitative and conceptual problem solving. They engage students actively in the learning process and
provide feedback based upon their needs.
They focus on helping students learn problem solving and reasoning skills while helping them build a more
coherent knowledge structure related to physics.
They can be used as a self-study tool by students.
The paired problems can be incorporated into regular quizzes or assigned as homework problems.

\begin{theacknowledgments}
We thank Daniel Haileselassie and Josh Bilak for help in the development and evaluation
of the tutorials and thank F. Reif, J. Levy, and R. Devaty for helpful discussions. 
We are grateful to the NSF for award DUE-0442087.
\end{theacknowledgments}

\bibliographystyle{aipproc}

\end{document}